\documentstyle[aps,preprint,floats,epsfig]{revtex}
\input epsf

\def\simlt{\stackrel{<}{{}_\sim}}

\def\btabl{\begin{table}}   \def\etabl{\end{table}}
\def\bea{\begin{eqnarray}}   \def\eea{\end{eqnarray}}
\def\bnn{\begin{eqnarray*}}   \def\enn{\end{eqnarray*}}
\def\beq{\begin{equation}}   \def\eeq{\end{equation}}  
\def\btabu{\begin{tabular}}   \def\etabu{\end{tabular}}
\def\bec{\begin{displaymath}} \def\eec{\end{displaymath}}

\def\eqref#1{(\ref{#1})}

\renewcommand{\baselinestretch}{1.2}
\begin{document}
\newcommand{\bfig}{\begin{center}\begin{picture}}
\newcommand{\efig}[1]{\end{picture}\\{\small #1}\end{center}}
\newcommand{\flin}[2]{\ArrowLine(#1)(#2)}
\newcommand{\wlin}[2]{\DashLine(#1)(#2){3}}
\newcommand{\zlin}[2]{\DashLine(#1)(#2){5}}
\newcommand{\glin}[3]{\Photon(#1)(#2){2}{#3}}
\newcommand{\lin}[2]{\Line(#1)(#2)}
\newcommand{\sof}{\SetOffset}
\draft

\preprint{\vbox{\baselineskip=15pt
\rightline{CERN-TH/99-299}
\rightline{LPT Orsay/00-60}
\rightline{hep-ph/0007041}}}
\title{Constraints on both  bilinear and trilinear R-parity violating 
couplings from neutrino
laboratories and astrophysics
data}
\author{
A. Abada$^{a}$\footnote{
e-mail: abada@th.u-psud.fr} and
M. Losada$^{b}$\footnote{
e-mail: malosada@venus.uanarino.edu.co}}
\vskip -0.5cm
\vspace{-0.5cm}
\address{{\small{\em $^a$ Laboratoire de Physique Th\'eorique\\
Universit\'e de Paris XI, B\^atiment 210, 91405 Orsay Cedex,
France\\
$^{b}$ Centro de Investigaciones, Universidad Antonio Nari\~{n}o,\\
Calle 59 No. 37-71, Santaf\'e de Bogot\'a, Colombia}}}

\vskip 3cm 

\maketitle \begin{abstract}

We consider neutrino masses generated at
tree level and at one loop, through fermion--sfermion 
loop diagrams, in the MSSM with
R-parity  violation. Using  the $(3\times 3)$ mass  and mixing
matrices 
for three generations  
of neutrinos and
the present experimental results on neutrinos from
laboratories and astrophysics  simultaneously, 
 we  put 
 bounds on both trilinear 
 ($\lambda_{ijk}, \lambda'_{ijk}$) and bilinear ($\mu_{e},\mu_\mu,\mu_\tau$)
  R-parity-violating couplings.

Pacs numbers:  13.15.+g, 14.70.Bh,  95.30.Cq.
\end{abstract}

\vspace{1.5cm}
 
\leftline{CERN-TH/99-299}
\leftline{June 2000}  
\renewcommand{\baselinestretch}{1.4}

\newpage \section{Introduction}

The analysis of the flavour structure of the neutrino mass matrix has been
a subject of intensive study recently 
\cite{neutrinos}.
 A non-trivial neutrino mass matrix
can solve, through the oscillation solutions, the atmospheric 
and solar neutrino
anomalies, which  have been observed by different experiments: to the very recent results of the Super-Kamiokande 
 collaboration \cite{ska},  have to be added  
those of  other
 atmospheric neutrino  experiments (IMB \cite{imb}, Soudan \cite{soudan},
Kamiokande \cite{ka}) and solar neutrino experiments (Homestake \cite{davis}, Gallex 
\cite{gallex}, SAGE \cite{sage}, Kamiokande \cite{ks}, 
Super-Kamiokande \cite{sks}, MACRO 
 \cite{macro} and LSND 
\cite{lsnd}).

 The oscillation explanation
 of the solar neutrino problem, the atmospheric
neutrino anomaly and the LSND results suggest three very different values of 
 neutrino mass squared
differences, namely $\Delta m^2_{{\mathrm{sun}}} \ll\Delta m^2_{{\mathrm{atm}}}
 \ll\Delta m^2_{{\mathrm{LSND}}}$,  with $\Delta m^2_{{\mathrm{sun}}} \simlt 10^{-4}$
 eV$^2$, 
 $\Delta m^2_{{\mathrm{atm}}}\in [10^{-3},10^{-2}]$ eV$^2$ and 
 $\Delta m^2_{{\mathrm{LSND}}}\in [0.3,1.0]$ eV$^2$.  The evidence for
 $\bar\nu_\mu\to\bar\nu_e$ observed by LSND has not been confirmed or excluded
 by the KARMEN \cite{karmen} experiment. MiniBooNE at FNAL \cite{Mboone} or MINOS long-baseline 
 experiments \cite{MINOS} 
 could provide  the answer. Excluding LSND  results, the oscillation explanation
 of the solar neutrino problem and the atmospheric
neutrino anomaly 
 requires two  mixing angles and
 two  values of 
 neutrino mass squared
differences with a strong hierarchy, namely $\Delta m^2_{{\mathrm{sun}}} \ll\Delta m^2_{{\mathrm{atm}}}$.

In a  previous  analysis \cite{abada-losada1} with a general $(3\times 3)$  symmetric mass matrix  assuming
 oscillations as an explanation of the neutrino deficits, we  
 had used as
  inputs    neutrino data constraints in order to obtain  the
allowed range of variation of
  the mass matrix elements. These inputs   are $\Delta
 m^2_{{\mathrm{atm}}}, \sin^2 2\theta_{{\mathrm{atm}}}$ from  SuperK
 measurements, $\Delta
 m^2_{{\mathrm{sun}}}, \sin^2 2\theta_{{\mathrm{sun}}}$ from solar
 experiments and finally the CHOOZ constraint \cite{chooz} 
  $\sin^2 \theta_{{\mathrm{CHOOZ}}}$. 
We also discussed  the spectrum after diagonalization and  considered two interesting quantities: 
the sum of the neutrino masses,
which is a relevant quantity for 
 obtaining a neutrino component of
hot dark matter (HDM), and 
the effective mass constraint, which is relevant for  neutrinoless double beta decay
(this is applicable  when the neutrinos are
 Majorana and massive particles).

To generate  a Majorana  mass for the neutrinos in a given  
model lepton number must be violated. 
We consider the Minimal Supersymmetric Standard Model (MSSM) with R-parity 
\cite{fayet} 
 violation
(${\mathrm{R}}_{\mathrm{P}}=(-1)^{L+3B+2S}$, where $L$, $B$, $S$ are the 
  lepton and 
 baryon number and the spin of the particle, respectively),   left-handed 
 neutrinos  obtain a
 Majorana mass, at tree level, through mixing with the neutralinos, 
 and through loop diagrams that violate lepton number (in two units). The simultaneous presence
of baryon and lepton number violating couplings is not
acceptable, because of the long lifetime of the proton. As we are interested in a model
that provides lepton-number violation, we simply choose all baryon number violating couplings
to be exactly zero.

 In ref. \cite{abada-losada1} we presented as  an application
of our results  the case of the Minimal Supersymmetric Standard Model (MSSM), with R-parity violation at the one-loop
order allowing the presence of  both bilinear and trilinear 
 ${\mathrm{R}}_{\mathrm{P}}$-violating couplings. We included the effect of
the bilinear couplings at tree level and of trilinear couplings
at one loop. The numerical bounds were given assuming 
all trilinear $\lambda_{ijk} \equiv \lambda$, as well as $\lambda'_{ijk} \equiv \lambda'$ to be equal. 
Here we will relax this assumption
and apply our general results to  other limiting (albeit general) cases in this model. The aim is
to provide strong constraints on the couplings with an explicit reference to the
generation indices from neutrino data. We will allow the simultaneous
presence of various lepton number violating couplings.

The paper is organized as follows. In section II we describe the inputs used
to constrain the R-parity violating parameters, while in section III
we introduce our notation
 for  
the MSSM with ${\mathrm{R}}_{\mathrm{P}}$-violation. 
   We also give in section III 
   the tree-level neutrino ($3\times 3$) mass matrix and the loop corrections 
to each matrix element. We parametrize the mass and mixing matrices.
  In section IV, we give results for  five separate subcases   
that correspond to some limiting cases of the MSSM 
without R-parity,
and  we
 derive  bounds on the bilinear and trilinear ${\mathrm{R}}_{\mathrm{P}}$-violating
  couplings for each case.

\section{General inputs}

 Global fits of neutrino data have shown that neutrino oscillations among three  
 flavours are sufficient to accommodate the solar and atmospheric data 
 \cite{Fogli--Bahcall}. 
 Three different types of solar experiments 
 are sensitive to different solar neutrino energy ranges; consequently, 
 three different ranges  of solutions for the solar data exist, 
   which correspond to the 
 vacuum
 oscillation solution, MSW with a large mixing angle (MSW-LMA) 
 and MSW with a small mixing angle (MSW-SMA).
The required
 neutrino mass squared differences
 and  mixing angles are shown in table I. 
We use these inputs together with the results  from  the CHOOZ experiment
 \cite{chooz}
to constrain the elements of our neutrino mass matrix in
the flavour basis. An additional  sixth input  is needed to solve the general ($3\times 3$)
real and symmetric mass matrix.
As our sixth input we could take 
  the direct upper bound on the effective mass from neutrinoless beta decay
 $(\beta\beta)_{0\nu}$,  
 or the upper bound on the effective  neutrino mass coming from 
the $^3 {\mathrm{H}}\ \beta$-decay spectrum, or the 
astrophysical bound on the magnetic moment of the neutrino bounds, 
 or the global cosmological upper bound on the sum of the
 neutrino masses 
 \cite{cosmo}. 

Let us briefly discuss each  of these possible inputs separately.
The direct upper bound  from the measurements of the high energy part
of the $^3 {\mathrm{H}}\ \beta$-decay spectrum given as upper limits on the 
electron neutrino mass obtained in
the TROITSK \cite{troitsk} and MAINZ \cite{mainz} experiments are 
$m_\nu < 2.5$ eV and  $m_\nu <
2.8$ eV, respectively.
 However, these experiments suffer from some ambiguities referred
 to as ``the
negative mass squared problem", which is still not completely  
understood and we will therefore not use these bounds.
 
 The upper bound on the sum of the  neutrino masses coming from astrophysical
 and cosmological considerations, such as the one from hot dark matter (HDM),
   suggests that 
 $$\sum_i m_{\nu_i} = m_{1} + m_{2} + m_{3}< {\mathrm few  \ \ eV}, $$
 where $m_i, i=1,2,3$, are the masses of the mass eigenstates ${\nu_i}$, which constitute active
 flavour neutrinos.
  A recent analysis, based on the observation of
 distant objects favouring a non-zero cosmological constant instead of HDM 
  \cite{cosmological}, shows that the bound of a few eV for the sum of the
neutrino masses is no longer required, since a 
 HDM component is not a necessary ingredient in this case.

In the R-parity violating model that we are considering we can
have transition magnetic moments that change lepton number. These
magnetic moments are proportional to the matrix elements; we prefer to use 
as an input a direct bound on a matrix element as such arising from neutrinoless double beta decay.

The  additional  input we will take  is  the best limit on the effective mass $m_{\mathrm{eff}}$ appearing in  
$(\beta\beta)_{0\nu}$, as it
directly constrains the matrix element $m_{11}$ of the neutrino
mass matrix. It is 
   defined by
\bea\label{beta-0}
|m_{\mathrm{eff}}|=
 |\sum_i m_{\nu_i} U_{e i}^2|\le \sum_k  m_{\nu_k}| U_{e k}^2|,
\eea
which  has been derived in the Heidelberg--Moscow $^{76}$Ge experiment 
\cite{G-G} (see also references
\cite{{M-H},{M-H-2}}): 
 \bea |m_{\mathrm{eff}}|\ < \  (0.2 ~- ~ 0.6)\ \ \mathrm{eV} \ \ 
 (\mathrm{at}\  90\%   \mathrm{CL}).\eea
The CP phases that might appear in the mixing matrix elements $U_{e i}$ are not relevant
 to our analysis, as we only use the second r.h.s. term in 
eq. (\ref{beta-0}).

For three generations, 
the flavour states $\nu_l$  are expressed in terms of the  mass 
eigenstates $\nu_i$  using the ($3\times3$)
 mixing matrix $U $
\bea
\nu_\ell= \sum_{i=1}^3 U^*_{\ell i} \nu_i.
\eea
Using  the Chau and 
 Keung parametrization of a $3\times
 3$ rotation matrix \cite{C-K}, $U$ is given by

 \bea\label{mix}
U =  \left(
 \begin{array}{ccc}
 c_{12}\ c_{13}&s_{12}\ c_{13}& s_{13} e^{-i \delta}\\
 -s_{12}\ c_{23}-c_{12}\ s_{23}\ s_{13}e^{i \delta}
 &c_{12}\ c_{23}-s_{12}\ s_{23}\
 s_{13}e^{i \delta}&s_{23}\ c_{13}\\
 s_{12}\ s_{23}-c_{12}\ c_{23}\ s_{13}e^{i \delta}
 &-c_{12}\ s_{23}-s_{12}\ c_{23}\ s_{13}e^{i \delta}&
 c_{23}\ c_{13}
 \end{array}
 \right)\  {\mathrm{diag}}\left\{e^{i \alpha_1},e^{i \alpha_2} ,1\right\},
 \eea
where $c_{ij}=\cos (\theta_{ij})$ and  $s_{ij}=\sin (\theta_{ij})$, $\delta$ is the
Dirac CP phase and $\alpha_{1,2}$ are the Majorana ones 
(we have two additional CP phases in the case of Majorana particles).

  The survival 
 probability $P_{\nu_e \to\nu_e}$, relevant to the case of solar fluxes, depends
 only on the first row of the mixing matrix in eq. (\ref{mix}), i.e. on 
 $|U_{ei}|^2$, with $i=1,2,3$. In the atmospheric case, the oscillation probability
 depends on the last column of (\ref{mix}), i.e. on 
 $|U_{\ell 3}|^2$, with $\ell=e,\mu,\tau$.
 The other elements of the matrix are not constrained by any direct experimental
 observation.
With this parametrization, we  directly obtain  the  mixing angles  $\sin^2 2 \theta_{12}$ and 
$\sin^2 2 \theta_{23}$, and also the relevant CHOOZ parameter $\sin\theta_{13}$.

 We impose the following hierarchy on  the mass eigenvalues and denote them by  $m_i, \ i=1,2,3$, 
 such that $m_1\leq m_2\leq m_3$. 
As the neutrino oscillation scenario cannot fix the absolute mass scale nor 
 distinguish whether the smallest mass splitting is between the two lightest mass 
 eigenstates or the two heaviest ones,
we present in table II  the possible neutrino spectra and indicate which is the corresponding  
mass squared difference.

To summarize, we will use the  following six inputs from neutrino data: 
\begin{itemize}
\item the limit  from the effective mass appearing in neutrinoless
double beta decay: $m_{\mathrm{eff}}$,
\item $\Delta m^2_{\mathrm{atm}}$ and $\sin ^2 2\theta_{\mathrm{atm}}$,
\item  $\Delta m^2_{\mathrm{sun}}$ and $\sin ^2 2\theta_{\mathrm{sun}}$,
\item $\sin^2 2\theta_{{\mathrm{CHOOZ}}}$.
\end{itemize}

\begin{table}[hbt]
\begin{tabular}{|c|l|l|}
Experiment & $\Delta m^2$ (eV$^2$) & $\sin^2 2\theta$\\\hline
Atmospheric & $(2-5)\times 10^{-3}$& $0.88-1$\\\hline
Solar&&\\
 MSW-{LMA}&$(3-30)\times 10^{-5}$ & $0.6\ -\ 1$\\
 MSW-{SMA}&$(0.4-1)\times 10^{-5}$ & $10^{-3}\ -\ 10^{-2}$\\
 Vacuum&$(0.5-8)\times 10^{-10}$ & $0.5\ -\ 1$\\\hline
CHOOZ & $> 3 \times 10^{-3}$& $<0.22$ \protect\label{fitss}
\end{tabular}
\vskip 0.5cm
\caption
{MSW-LMA, MSW-SMA and Vacuum stand for MSW large mixing angle, small mixing and vacuum 
oscillation solutions, 
respectively.}
\end{table}

\vskip 1cm

\begin{table}[hbt]
\begin{tabular}{|c|l|l|}
Spectrum  &Solar& Atmospheric \\ \hline
Hierarchy  &$\Delta m_{12}^2$& $\Delta m_{13}^{2}$ \\ \hline
Degenerate & $\Delta m_{23}^2 \quad{\mathrm{or}}\quad \Delta m_{12}^{2}$ & 
$\Delta m_{13}^{2}$ \\ \hline
Pseudo-Dirac & $\Delta m_{23}^2 \quad{\mathrm{or}}\quad \Delta m_{12}^{2}$ & 
$\Delta m_{13}^{2}$ \protect\label{spectra}
\end{tabular}

\vskip 0.5cm
\caption{Different possible regimes and corresponding  mass squared
difference. }
\end{table}

\newpage
 \section{MSSM with  ${\mathrm{R}}_{\mathrm{P}}$- violation}

The most general renormalizable superpotential for the supersymmetric
 Standard Model with lepton-number violation is

\bea
W = \epsilon_{ab}[\mu_{\alpha} \hat{L}_{\alpha}^{a} \hat{H}_{u}^{b}} + \lambda_{\alpha\beta k}\hat{L_{\alpha}^{a} \hat{L}_{\beta}^{b}\hat{E}_{k}^{C}  + h_{ik}^{u} \hat{Q}_{i}^{a}
\hat{H}_{u}^{b} \hat{U}_{k}^{C}
 + \lambda'_{\alpha i k} \hat{L}_{\alpha}^{b} \hat{Q}_{i}^{a} \hat{D}_{k}^{C}],
\label{W}
\eea
where the $(i,j,k)$ are flavour indices, $(a,b)$ are $SU(2)$ indices, and the $(\alpha,\beta)$ are
flavour indices running from 0 to 3. The $\hat{L}_\alpha$ are the doublet superfields
with hypercharge $Y=-1$. Note that the $\lambda$ couplings are antisymmetric in the first two
indices. The usual R-parity-preserving Lagrangian is obtained when only 
$\mu_{o}, \lambda_{0ik} = h_{ik}^{d}, \lambda'_{i0k}= h_{ik}^{d}$ are non-zero, 
and we can identify $\hat{L}_{o} \equiv \hat{H}_{d}$.
In the model of eq. (\ref{W}) we have 9 additional $\lambda$ couplings and 27 new $\lambda'$ 
couplings with respect
to the R-parity-conserving case. 
Note that thanks to the additional degrees of freedom, we can rotate in the flavour space of
the ``down-type'' scalar fields to set the vacuum expectation values 
of the sneutrinos to be zero~\footnote{This can be done order by order 
in the loop expansion when one appropriately defines the mass matrices of
the Higgs sector.}. Henceforth, we will work consistently in this basis and the bounds we will derive are valid for this basis.

This model has been 
extensively analysed in the literature  
 \cite{barbieri}-- \cite{banks}.
 References \cite{nardi,banks} showed 
in a basis-invariant way
that neutrino masses are always generated in these models, even when universality of
the soft SUSY-breaking terms is assumed at some high scale. 
Previous studies have also tried to constrain the different
 ${\mathrm{R}}_{\mathrm{P}}$-violating couplings that 
 appear in the MSSM Lagrangian,  considering only the effect of bilinear terms  
\cite{kaplan} 
or only of trilinear couplings \cite{choi},  
 or of  both 
\cite{chunetal,abada-losada1}, 
 from  solar and atmospheric neutrino data.
Both tree-level and one-loop effects have been considered.
A recent study \cite{ChunChoi} has constrained these couplings
using rare decays. The results are in agreement with the constraints
given in ref. \cite{abada-losada1}.
\begin{figure}[hbt]
\vspace{40pt}
\hspace{1.3cm}\epsfxsize=6cm\epsfbox{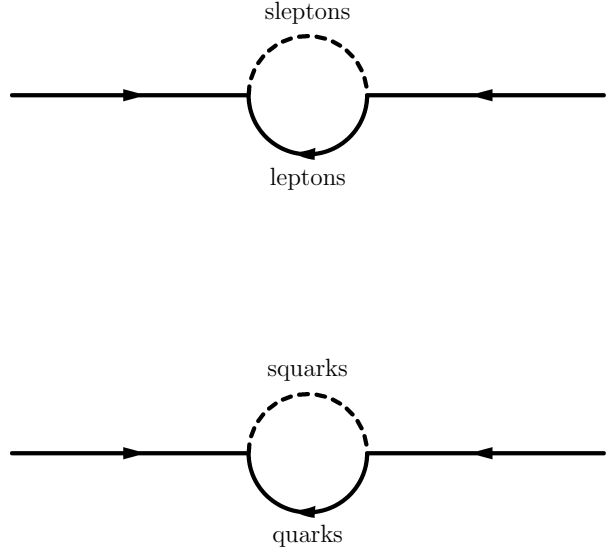}
\vspace*{-2.5cm}\vskip -4cm
\hspace{3.3cm}\caption[]{\it{One-loop Feynman diagrams contributing to the neutrino masses.}} 
 \protect\label{fig1}
\end{figure}

It is well known that the tree-level expression of the neutrino
mass matrix elements obtained   through the mixing with neutralinos is
\begin{equation}
M^{\mathrm{tree}}_{\nu_{ij}} = g_{2}^{2}{(M_{1} +
 \tan^2\theta_{W} M_{2})\over 4 
\det M}\mu_{i}\mu_{j} v_{1}^{2} \hspace{0.2cm} \sim g_2^2 
{v^2 \cos^2\beta\over M_{\mathrm{susy}}^3}\mu_i \mu_j\equiv C \mu_i \mu_j ,
\label{treeelem}
\end{equation}
where $i,~j$ are flavour indices, $M_1, M_2$ are the gaugino masses and $\det M$ is the determinant
of the R-parity conserving neutralino mass matrix. For the second term
of the right-hand side we have taken all R-parity conserving masses to be
of order $\sim M_{\mathrm{susy}}$ and as we are working in the basis where the slepton vacuum 
expectation value is zero, $v_{1} = v_{d} = v \cos\beta$. We can rewrite
the tree-level mass matrix as  
 \bea\hspace*{-0.5cm} {\cal{M}}^{\mathrm{tree}}_{\nu}= 
 C \left(
 \begin{array}{ccc}
 \mu_e^2& \mu_\mu\mu_e& \mu_e\mu_\tau \\
\mu_\mu\mu_e& \mu_\mu^2& \mu_\mu\mu_\tau \\
 \mu_e\mu_\tau& \mu_\mu\mu_\tau& \mu_\tau^2
  \end{array}
 \right)\ .
\label{MRptree}
 \eea

The above mass matrix has only one non-zero eigenvalue.  In order for the other neutrinos to obtain a mass, we must
consider loop corrections. There are
many types of loop diagrams that can  contribute to the
neutrino mass in this model \cite{Haber,Kong2,dl}. Here we will focus on the
contribution at one loop from fermion--sfermion diagrams
 involving trilinear couplings, see fig. \ref{fig1}. See, for example,
 ref.
\cite{abada-losada1} for more details regarding the computation of
the neutrino mass matrix.

The one-loop mass contribution
from slepton--lepton and squark--quark
 loops to each element
of the mass matrix is given by

\bea
(m_{qm})_{\mathrm{loop}} = {1\over 16 \pi^2} X\left({f(x_\ell) \over M^{2}_2} 
\sum_{k,p} \lambda_{qkp}\lambda_{mpk} m_{\ell}^{(k)}m_{\ell}^{(p)} + 3{f(x_q) \over M_{q2}^2}
 \sum_{k,p} \lambda'_{qkp}\lambda'_{mpk} m_{q}^{(k)} m_{q}^{(p)}\right),
\label{loopmnew}
\eea
where
\bea
f(x)=-{\ln \ x\over 1-x},\quad
x_\ell^{(p)}= \left({M_1^{(p)}\over M_2^{(p)}}\right)^2\quad, \quad
x_q^{(p)}= \left({M_{q1}^{(p)}\over M_{q2}^{(p)}}\right)^2  {\mathrm{and}}\quad X= A+ \mu\tan\beta,
\eea
where $X$ is the trilinear term that appears in the off-diagonal
matrix element of the sfermions mass matrix.
We consider here that we are in the down-quark mass eigenstate basis, and that the
$\lambda,\lambda'$ couplings have been redefined in terms of the couplings
appearing in the superpotential and of the corresponding Cabibbo--Kobayashi--Maskawa
matrix elements; we also drop the flavour dependence from 
$ { X\over M^{(p)\ 2}_2} f(x_\ell^{(p)})$ and ${ X\over M^{(p)\
2}_{q2}} f(x_q^{(p)})$, and consider them to be universal in the slepton and
in the squark sector, respectively. Thus the matrix elements at one loop are
given by the sum of eqs. (\ref{MRptree}) and (\ref{loopmnew}).
 The analysis we perform can directly  constrain  terms of the form $C \mu_i \mu_j$,  $K_1^{ij} \sum_{k,p} \lambda_{ikp}\lambda_{jpk}$, $K_2^{ij} \sum_{k,p} \lambda'_{ikp}\lambda'_{jpk}$, where

\begin{eqnarray}
K_1^{ij} & =& {X\over 16 \pi^2} {f(x_\ell) \over M^{2}_2} \left(
  m_i m_j  \right)\ , \nonumber \\
K_2^{ij}&=&3  {X\over 16 \pi^2} {f(x_q) \over M^{q2}_2} \left(  m_i m_j \right)\ .
\label{k1k2ij}
\end{eqnarray}
Note that the quantities $K^{ij}_{1,2} $  depend only on
R-parity conserving parameters.
As an application, we focus in the following on a case where  
we assume that the $\lambda'_{ijk} $ are of the same order, which allows
us to neglect the extra mixed  $\lambda'_{123,132,223,232} $ due to the
hierarchy $m_sm_b\ll m_b^2$ (see eq. (\ref{loopmnew})). 
We also assume separately  that all $\lambda_{ijk}$
are of the same order~\footnote{The same procedure as we apply here could be performed if we considered   that only one of the R-parity violating trilinear couplings
was dominant.}.
Thus, using the  mass hierarchy 
$m_{e,d}\ll m_{\mu,s}\ll m_{\tau,b}$ in $(m_{qm})_{\mathrm{loop}}$, 
simplifies the expressions for the matrix elements, to obtain

 \bea\hspace*{-0.5cm} 
{\cal{M}}^{\mathrm{loop}}_{\nu}= \left(
 \begin{array}{lll}
 K_1\lambda^2_{133}+K_2\lambda'^2_{133}& 
K_1\lambda_{133}\lambda_{233}+
K_2\lambda'_{133}\lambda'_{233} &
 K_2\lambda'_{133}\lambda'_{333}\\
 K_1\lambda_{133}\lambda_{233}+
K_2\lambda'_{133}\lambda'_{233}&  
 K_1\lambda^2_{233}+K_2\lambda'^2_{233}&  
 K_2\lambda'_{233}\lambda'_{333} \\ 
K_2\lambda'_{133}\lambda'_{333} &
K_2\lambda'_{233}\lambda'_{333}  
& K_2\lambda'^2_{333}
 \end{array}
 \right) ,
\label{MRploop}
 \eea
 where the coefficients $K_{1,2}$ are given by:
\begin{eqnarray}
K_1 & =& {X\over 16 \pi^2} {f(x_\ell) \over M^{2}_2} \left(
  m_{\tau}^2  \right)\ , \nonumber \\
K_2&=&3  {X\over 16 \pi^2} {f(x_q) \over M^{q2}_2} \left(  m_{b}^2 \right)\ .
\label{AAB}
\end{eqnarray}

Thus, we take our one-loop mass matrix to be
  \bea
  {\cal{M}_{\nu}}={\cal{M}}^{\mathrm{tree}}_{\nu}
+{\cal{M}}^{\mathrm{loop}}_{\nu}. 
\label{Moneloop}
\eea
At this order, the mass matrix has in general three non-zero eigenmasses. The
 {\bf 8} R-parity violating parameters that we would
like to constrain are:
\begin{itemize}
\item $\lambda_{133}, \lambda_{233}$  
\item $\lambda'_{133}, \lambda'_{233}$ and $\lambda'_{333}$

\item  $\mu_e$, $\mu_\mu$ and $\mu_\tau$.

\end{itemize}
Recall that we will use only {\bf 6} inputs from 
 solar (2), atmospheric (2), CHOOZ (1) constraints and the bound on the effective neutrino mass 
 appearing in neutrinoless double beta decay.  

In  our previous analysis \cite{abada-losada1}, denoted case 0 in table III, we  considered a toy model 
leading to three non-zero mass eigenstates. In this toy model we 
 assumed that
all  trilinear couplings $\lambda_{ijk}$ were equal
 and that all 
 $\lambda'_{ijk}$ couplings were equal, i.e. $\lambda_{ijk}=\lambda$ and
  $\lambda'_{ijk}=\lambda'$.
 This leads to five unknowns 
 $\mu_{e,\mu,\tau}$ , $\lambda$ and $\lambda'$, and the system was
 solved using the five inputs from the solar, atmospheric and  
 CHOOZ constraints. 

In this work, we will relax this assumption to more general cases. 
We consider five different cases
 that generate three non-zero  eigenmasses.
  These different subcases of the generic matrix of eq. (\ref{Moneloop}), 
arise when
we apply certain conditions on some of the $L$-number violating parameters. 
We do
this so as to reduce the number of unknowns  to the number of
constraints. The subcases we consider are summarized in table \ref{subcases}. 

A first case  corresponds to the situation where all the  $\lambda'$ couplings are
switched off, which happens in  the limit in which the squarks decouple.
Case 2 occurs when all the  $\lambda$ couplings are
switched off, i.e. in  the limit in which the sleptons decouple.
In  case 3, the bilinear couplings are such that 
$\mu_e\sim \mu_\mu \ll\mu_\tau$ and thus only one bilinear coupling 
($\mu_\tau$) is
relevant.  
In  cases 4 (5), the trilinear $\lambda$ ($\lambda'$) are of the same 
order, 
i.e. $\lambda_{133}\simeq \lambda_{233}= \lambda$ 
($\lambda'_{133}\simeq \lambda'_{233}\simeq \lambda'_{333}= \lambda'$) 
and the bilinears are such that
 $\mu_e\sim \mu_\mu \ne\mu_\tau$ for case 4 and $\mu_e\ne\mu_\mu \ne\mu_\tau$
 for case 5.
We do not claim
to have covered all possibilities, but certainly some of the most straightforward ones. 

\begin{table}[htb]
\begin{tabular}{|c|c|c|c|}
Cases  & $\lambda_{ijk}$&$\lambda'_{ijk}$ &$\mu_{i}$ \\ \hline
0  &$\lambda_{ijk} =\lambda$ &$\lambda'_{ijk} =\lambda'$ &$\mu_{e}, 
\mu_{\mu},\mu_{\tau}$ \\ \hline
1 & $\lambda_{133},\lambda_{233}$ & 0&$\mu_{e}, 
\mu_{\mu},\mu_{\tau}$ \\ \hline
2 &  0&$\lambda'_{133},\lambda'_{233},\lambda'_{333}$ &$\mu_{e}, 
\mu_{\mu},\mu_{\tau}$ \\ \hline
3 &$\lambda_{133},\lambda_{233}$ &$\lambda'_{133},\lambda'_{233},\lambda'_{333}$ &$\mu_{\tau}$ \\ \hline
4 &$\lambda_{ijk} =\lambda$ &$\lambda'_{133},\lambda'_{233},\lambda'_{333}$ &$\mu_{\tau}$,$\mu_e=\mu_{\mu}$ \\ \hline
5 & $\lambda_{133}, \lambda_{233}$ &$\lambda'_{ijk} =\lambda'$&$\mu_{e}, 
\mu_{\mu},\mu_{\tau}$ 
\protect\label{subcases}
\end{tabular}

\vskip 0.5cm 
\caption{Different cases of  ${\mathrm{R}}_{\mathrm{P}}$-violating couplings contributing to the
neutrino mass matrix.}
\end{table}
\newpage
\section{Results}

In  our general scan of parameter space, we  allow tree-level contributions to either dominate over the loop corrections,
or to be on the same order as these, or to be much smaller than the loop terms.
The analysis for subcase $0$, presented in 
\cite{abada-losada1}, gave bounds on the $R_{p}$ couplings
from combinations of constraints from atmospheric and CHOOZ data, together with
one of the possible solar neutrino solutions. For our five new subcases we 
 relax the assumptions of  subcase $0$ to present 
stringent constraints on bilinear and trilinear lepton-number violating couplings,
 with specific generation indices in the basis where the sneutrino vacuum expectation
value is zero. 

 The bounds on the couplings are presented in tables for the various combination of
constraints from neutrinoless double beta decay, atmospheric, CHOOZ together with vacuum or  MSW-SMA or MSW-LMA
solutions.
A common assumption used to place bounds in this model is to take  all 
${\mathrm{R}}_{\mathrm{P}}$-conserving
 mass parameters to be of the same order,
 $M_{\mathrm{susy}}$. For the particular case where $M_{\mathrm{susy}}=100$ GeV,
  $f(x) \rightarrow 1$ and $\tan\beta =2$, 
  we have 
  $$K_1 \sim 1.8\times 10^{-4}  {\mathrm{GeV}},\hspace{0.2cm}
K_2 \sim 4.7\times 10^{-3} {\mathrm{GeV}},\hspace{0.2cm} C \sim 5.3 \times 10^{-3}
{\mathrm{GeV}}^{-1}. $$ We  use these values to obtain
the numerical results presented for cases 1--5 in tables IV-VIII, respectively.
Using eqs. (\ref{treeelem}) and (\ref{AAB}) modified bounds can be obtained for other values of the
R-parity conserving parameters.

\begin{table}[htb]

\begin{tabular}{|c|c|c|c|}
Couplings & MSW-LMA & MSW-SMA& Vacuum\\\hline
$|\lambda_{133}|$ & $6.8\times 10^{-4}$  & $1.3\times 10^{-4}$&- \\\hline
$|\lambda_{233}|$ & $5.6\times 10^{-4}$& $1.9\times 10^{-4}$&- \\\hline
$|\mu_e|\ ({\mathrm{GeV}}$)& $6.1\times 10^{-5}$& $2.0\times 10^{-6}$&- \\\hline
$|\mu_\mu|\ ({\mathrm{GeV}}$)& $1.2\times 10^{-4}$& $1.2\times 10^{-4}$& -\\\hline
$|\mu_\tau|\ ({\mathrm{GeV}}$)& $1.2\times 10^{-4}$& $1.2\times 10^{-4}$&- 
\end{tabular}
\caption{Bounds on the couplings for subcase 1
 that satisfy MSW-LMA or MSW-SMA,
 SuperK, CHOOZ and neutrinoless double beta decay simultaneously.}
\end{table}
 \begin{table}[htb]
\begin{tabular}{|c|c|c|c|}
Couplings & MSW-LMA & MSW-SMA& Vacuum\\\hline
$|\lambda'_{133}|$ & $1.5\times 10^{-4}$  & $4.1\times 10^{-5}$&$5.4\times 10^{-5}$\\\hline
$|\lambda'_{233}|$ & $1.5\times 10^{-4}$  & $1.5\times 10^{-4}  $&$ 9.5\times 10^{-5}   $\\\hline
$|\lambda'_{333}|$ & $1.5\times 10^{-4}$  & $1.5\times 10^{-4}  $&$1.5\times 10^{-5}$\\\hline
$|\mu_e|\ ({\mathrm{GeV}}$)& $1.2 \times 10^{-4}$& $4.1\times 10^{-5}$&$5.1\times 10^{-5}$\\\hline
$|\mu_\mu|\ ({\mathrm{GeV}}$)& $1.2 \times 10^{-4}$& $1.2 \times 10^{-4}  $&$8.1\times 10^{-5}$\\\hline
$|\mu_\tau|\ ({\mathrm{GeV}}$)& $1.2\times 10^{-4}$& $ 1.2 \times 10^{-4}$&$1.2 \times 10^{-4}$
\end{tabular}
\caption{Bounds on the couplings for subcase 2
 that satisfy MSW-LMA or MSW-SMA or vacuum, SuperK, CHOOZ and the neutrinoless beta decay constraints
   simultaneously.}
\end{table}
\begin{table}[htb]
\begin{tabular}{|c|c|c|c|}
Couplings & MSW-LMA & MSW-SMA& Vacuum\\\hline
$|\lambda'_{133}|$ & $1.2\times 10^{-4}$  & $6.8\times 10^{-5}$&$1.4\times 10^{-5}$ \\\hline
$|\lambda'_{233}|$ & $1.5\times 10^{-4}$  & $1.5\times 10^{-4}$& $1.4\times 10^{-5}$\\\hline
$|\lambda'_{333}|$ & $1.5\times 10^{-4}$& $1.5\times 10^{-4}$&$1.5\times 10^{-4}$ \\\hline
$|\lambda_{133}|$ & $6.2\times 10^{-4}$  & $3.1\times 10^{-4}$&$1.1\times 10^{-7}$ \\\hline
$|\lambda_{233}|$ & $6.2\times 10^{-4}$& $3.1\times 10^{-4}$&$1.1\times 10^{-7}$ \\\hline
$|\mu_\tau|\ ({\mathrm{GeV}}$)& $1.3\times 10^{-4}$& $1.3\times 10^{-4}$&$1.2\times 10^{-5}$ 
\end{tabular}
\caption{Bounds on the couplings for subcase 3
 that satisfy MSW-LMA or MSW-SMA or vacuum, SuperK, CHOOZ and the neutrinoless beta decay constraints
   simultaneously.}
\end{table}
\begin{table}[htb]
\begin{tabular}{|c|c|c|c|}
Couplings & MSW-LMA & MSW-SMA& Vacuum\\\hline
$|\lambda'_{133}|$ & $1.1\times 10^{-4}$  & $3.4\times 10^{-5}$&$5.4\times 10^{-5}$ \\\hline
$|\lambda'_{233}|$ & $1.4\times 10^{-4}$  & $1.4\times 10^{-4}$& $5.4\times 10^{-5}$\\\hline
$|\lambda'_{333}|$ & $1.5\times 10^{-4}$& $1.4\times 10^{-4}$&$1.5\times 10^{-4}$ \\\hline
$|\lambda|$ & $4.4\times 10^{-4}$  & $1.9\times 10^{-4}$&$1.1\times 10^{-7}$ \\\hline
$|\mu_e\sim\mu_\mu |\ ({\mathrm{GeV}}$)& $8.1\times 10^{-5}$& $2.3\times
10^{-5}$&$4.6\times 10^{-5}$\\\hline
$|\mu_\tau|\ ({\mathrm{GeV}}$)& $1.2\times 10^{-4}$& $1.3\times 10^{-4}$&$1.3\times 10^{-4}$
\end{tabular}
\caption{Bounds on the couplings for subcase 4
 that satisfy MSW-LMA or MSW-SMA or vacuum, SuperK, CHOOZ and the neutrinoless beta decay constraints
   simultaneously.}
\end{table}

\begin{table}[htb]
\begin{tabular}{|c|c|c|c|}
Couplings & MSW-LMA & MSW-SMA& Vacuum\\\hline
$|\lambda_{133}|$ & $6.2\times 10^{-4}$  & $1.3\times 10^{-4}$&- \\\hline
$|\lambda_{233}|$ & $5.6\times 10^{-4}$  & $3.1\times 10^{-4}$&- \\\hline
$|\lambda'|$ & $6.8\times 10^{-5}$  & $4.1\times 10^{-5}$& -\\\hline
$|\mu_e|\ ({\mathrm{GeV}}$)& $8.1\times 10^{-5}$& $3.4\times 10^{-4}   $& -\\\hline
$|\mu_\mu|\ ({\mathrm{GeV}}$)& $1.3\times 10^{-4}$& $1.3\times 10^{-4}$&- \\\hline
$|\mu_\tau|\ ({\mathrm{GeV}}$)& $1.3\times 10^{-4}$& $1.3\times 10^{-4}$&- 
\end{tabular}
\caption{Bounds on the couplings for subcase 5
 that satisfy MSW-LMA or MSW-SMA or vacuum, SuperK, CHOOZ and the neutrinoless beta decay constraints
   simultaneously.}
\end{table}

The analysis for all subcases shows that there are many regions of parameter
 space that
can simultaneously  accommodate  the MSW solution (large or small mixing angle), SuperK and
CHOOZ constraints. Subcases 2--4 also provide solutions for the 
combined constraint of atmospheric data,
CHOOZ and the vacuum oscillation solution, while subcases 1 and 5 do not.
 The results for the latter two subcases are similar to those obtained for subcase 0 
treated in
\cite{abada-losada1}. The number of solutions defining the  allowed region in the $(\Delta m^2, \sin^2 2\theta)$  plane is very 
small for the
vacuum solution, larger in the case of MSW-SMA, and it is still larger in the case
of MSW-LMA. This is in agreement with 
the most recent results from the Super-Kamiokande collaboration, presented at the
 SUSY2K 
conference \cite{Totsuka}, stating that
  the solar neutrinos  favour the  MSW-LMA solution. There is less room for
 MSW-SMA and much less still for vacuum solutions. 
We can see that our bounds are consistent with 
bounds derived in the literature (see \cite{dreiner,bhatta2,bhatta3} for 
 three-neutrino case  and  \cite{bhatta4} for four-neutrino case). We emphasize that some of our subcases present  more stringent bounds, and that
the combination of constraints with the vacuum solution requires the smallest values for R-parity violating couplings.

To summarize, we have obtained bounds on the
bilinear and trilinear R-parity violating couplings
with explicit reference to the leptonic indices.
The bounds have been obtained from neutrino data,
which constrain the neutrino mass matrix that can
be constructed in the MSSM with R-parity violation
for three generations of neutrinos. We considered
the tree-level contribution and
the one-loop contribution from fermion--sfermion diagrams with trilinear
couplings  to the neutrino mass matrix.

\subsection*{Acknowledgements}
We thank G. Bhattacharyya for useful discussions. We also want to thank
the CERN TH Division for kind hospitality during the completion of this work.

\end{document}